# Active inference and artificial reasoning

Karl Friston[1,2], Lancelot Da Costa[2,3], Alexander Tschantz[2,4], Conor Heins[2,5], Christopher Buckley[2,4], Tim Verbelen[2], Thomas Parr[6]

[1] Queen Square Institute of Neurology, University College London, UK

[2] VERSES, Los Angeles, California, 90016, USA

[3] ELLIS Institute Tübingen, Tübingen, Germany

[4] Department of Informatics, University of Sussex, Brighton, UK

[5] Max Planck Institute of Animal Behavior, Department of Collective Behaviour, Konstanz Germany

[6] Nuffield Department of Clinical Neurosciences, University of Oxford, UK

Emails: k.friston@ucl.ac.uk; lance.dacosta@verses.ai; alec.tschantz@verses.ai; conor.heins@verses.ai; christopher.buckley@verses.ai; tim.verbelen@verses.ai; thomas.parr@ndcn.ox.ac.uk

**ABSTRACT**

This technical note considers the sampling of outcomes that provide the greatest amount of information about the structure of underlying world models. This generalisation furnishes a principled approach to structure learning under a plausible set of generative models or hypotheses. In active inference, policies—i.e., combinations of actions—are selected based on their expected free energy, which comprises expected information gain and value. Information gain corresponds to the KL divergence between predictive posteriors with, and without, the consequences of action. Posteriors over models can be evaluated quickly and efficiently using Bayesian Model Reduction, based upon accumulated posterior beliefs about model parameters. The ensuing information gain can then be used to select actions that disambiguate among alternative models, in the spirit of optimal experimental design. We illustrate this kind of active selection or reasoning using partially observed discrete models; namely, a 'three-ball' paradigm used previously to describe artificial insight and 'aha moments' via (synthetic) introspection or sleep. We focus on the sample efficiency afforded by seeking outcomes that resolve the greatest uncertainty about the world model, under which outcomes are generated.

# INTRODUCTION

Active inference offers a first principles approach to sense and decision-making under generative or world models. It distinguishes itself from (model-free) reinforcement learning—and related approaches in machine learning—in several respects. First, it draws a clear distinction between the unknown variables entailed by a generative model; namely, latent *states*, *parameters* and *structure*, c.f., (Kemp and Tenenbaum, 2008) This calls for inference over the three kinds of unknowns, leading to active *inference*, *learning* and *selection*, respectively (Parr et al., 2022). Crucially, latent states can include an agent's action or policies, leading *planning* as active inference, c.f., (Attias, 2003; Botvinick and Toussaint, 2012). This speaks to another distinction between active inference and learning state-action policies; in the sense that planning rests on free energy *functionals* of beliefs about states, as opposed to value *functions* of states *per se*. This means that active inference optimises (Bayesian) beliefs about latent variables. A key consequence is that an optimal policy resolves uncertainty—an attribute of a Bayesian belief—through exploratory, information-seeking actions that cannot be cast as optimising a (value) function of states *per se*.

The ensuing information-seeking is underwritten by the information gain expected under any given combination of actions, i.e., policy. The expected information gain over states is often referred to in terms of salience, corresponding to notions of Bayesian surprise in the visual search literature and accompanying salience maps (Itti and Baldi, 2009; Parr and Friston, 2017). In robotics, this kind of motivation is sometimes referred to as intrinsic—c.f., (Oudeyer and Kaplan, 2007)—that manifests as artificial curiosity (Schmidhuber, 2010). When applied to the parameters of a generative model, the expected information gain is sometimes referred to as novelty; namely, the epistemic affordance offered by resolving uncertainty about the parameters of a generative model (Barto et al., 2013; Schwartenbeck et al., 2019). In this paper, we introduce a third sort of information gain that pertains to the structure of the generative model (Da Costa et al., 2025; Tsividis et al., 2021). Equipping agents with this kind of intrinsic motivation is in line with the principles of optimum Bayesian design (Lindley, 1956; Mackay, 1992; Parr et al., 2024); namely, designing an experiment to elicit data that maximally disambiguate hypotheses or generative models.

The ability to select observations that minimise uncertainty or ambiguity is a hallmark of natural intelligence and self-organisation of certain natural kinds (Friston et al., 2023). One could argue that such behaviours rest upon inference—i.e., variational free energy minimising processes—that distinguish between artificial intelligence and natural intelligence; sometimes cast in terms of System one versus System two thinking, e.g., (Kelly and Barron, 2022). One might further argue that the homologue of 'thinking' in active inference is the planning (as inference) in which counterfactual futures are evaluated under plausible policies. This evaluation is in terms of the (path integral of) expected free energy that comprises expected information gain and expected value (Friston et al., 2023). Value—in the setting of active inference—corresponds to the log probability of *a priori* preferred outcomes. Expected information gain or epistemic value scores the degree of belief updating an agent expects under each policy. It is the expected information gain that is beyond the reach of conventional learning schemes because the information gain depends upon posterior beliefs that, by definition, are time, context, and experience dependent. In other words, there is no learnable mapping from observations to expected free energy—and implicitly policies—because the mapping *per se* changes with learning. In short, although one can learn to infer, e.g., (Mazzaglia et al., 2022), one cannot learn to be curious. The So, does planning with epistemic intentions provide a sufficient account of reasoning?

One would probably argue no. Reasoning—especially as used in the current large language model literature (Xu et al., 2025)—implies a degree of recursive introspection that, in the active inference literature, is better considered in terms of Bayesian model reduction. Bayesian model reduction is a fast and efficient form of Bayesian model selection that allows one to evaluate the evidence (i.e., marginal likelihood) of alternative models based upon the prior and posterior of some full or parent model (Dickey, 1971; Friston et al., 2018). It is usually implemented offline as a variational free energy minimising process of the sort that has been proposed to occur during introspection or sleep (Friston et al., 2017; Hinton et al., 1995).

Technically, variational free energy is model complexity minus accuracy, where complexity is the KL divergence between the posterior and prior (Penny, 2012). Active inference and learning therefore seek the posterior that minimises the divergence from prior beliefs, while providing an accurate account of observed data. However, after all the data have been observed, one can minimise variational free energy by selecting a prior that minimises the divergence from the posterior. This *post-hoc* selection is generally from a set of priors—over model parameters—that constitute a space of models or hypotheses. One can regard this form of selection as the basis of the scientific process in the following sense: a scientist elaborates a small model space (e.g., an alternate and null hypothesis), acquires some data under uninformative priors over her model space, and then finds the hypothesis that best explains her experimental data (Corcoran et al., 2023; Lindley, 1956). On this view, the best model is that which renders the data the most likely; namely maximises model evidence or marginal likelihood.

One can generalise this *post-hoc* variational free energy minimising process to any number of hypotheses—or reduced models—that can be specified in terms of priors over model parameters. Having identified the best hypothesis or model—following some suitable period of evidence accumulation—an agent can proceed with an informed and optimal model structure. In what follows, we will refer to this form of structure learning (Gershman and Niv, 2010; Tenenbaum et al., 2011) as applying *Occam's razor*. This application of Bayesian model reduction has been used to simulate 'aha moments' and 'abstract reasoning' using a three-ball paradigm, in which (synthetic) subjects had to discover the rule underlying 'correct' choices (Friston et al., 2017). Here, we reprise this paradigm to demonstrate active reasoning or reduction, where the focus is on selecting the right data—c.f., performing the right experiments—that resolve uncertainty about the models or hypotheses entertained. For the purposes of this paper, we will read 'reasoning' as selecting the best explanation for observations under competing hypotheses about how observations are generated. Active reasoning can then be read as seeking observations that best disambiguate competing hypotheses for subsequent 'reasoning'.

We will see that the requisite expected information gain can be computed quickly and efficiently—using Bayesian model reduction—and used to supplement the expected free energy that subtends action selection. This extension of active inference, to cover abstraction or reasoning (Chollet, 2019), was inspired largely by the ARC-AGI-3 challenge ([ARC-AGI-3](ARC-AGI-3)); namely, the challenge to discover a rule, symmetry, invariance, law, or abstraction for which no prior information is available. The procedures described in this paper address part of this challenge by describing an efficient way of querying or acting upon the world to discover the rules or contingencies in play. This is only a partial solution to the larger challenge, in that we assume that the agent has already learned a generative model of the puzzle or game at hand but does not know what constitutes a correct response. However, knowing the latent cause-effect structure of the game admits a set of hypotheses or models that conform to some symmetry constraints. In particular, we will consider hypothetical likelihood mappings from latent causes to successful outcomes that feature certain isomorphisms. This allows the agent to successively eliminate unlikely mappings—through active reduction or reasoning—until the rule is discovered. The resulting scheme shares key commitments with probabilistic inductive logic programming (Lake et al., 2015; Riguzzi et al., 2014); especially, the distinction between parameter and structure learning. However, its focus is on the sample efficiency of the implicit inductive reasoning.

This paper has three sections. The first provides a brief rehearsal of belief updating and propagation under the partially observed Markov decision processes used in subsequent numerical studies. This section focuses on the expected information gain over models that induces active reasoning. And how this component of expected free energy can be evaluated simply and automatically using Bayesian model reduction. The ensuing scheme is then applied to the three-ball paradigm described in the second section. This paradigm has several features that call for the epistemic foraging. In particular, the problem is partially observed; in the sense that the agent can only see the colour of each ball by selecting it (e.g., foveating the location of each ball). In turn, this calls for sequential policy optimisation that would elude conventional (e.g., reinforcement learning) schemes that seek state-action policies. This section simulates Bayes-optimal behaviour, in the dual sense of optimal Bayesian decision theory and experimental design. The third section repeats the simulations with and without various components of expected free energy to characterise their contribution to sample efficiency and performance. The paper concludes with a brief discussion of what is, and is not, covered by this application of active inference.

# ACTIVE INFERENCE, LEARNING AND SELECTION

Active inference rests upon a *generative model* of observable outcomes. This model is used to infer the most likely causes of outcomes in terms of expected states of the world. These states—and their transition or paths—are latent or *hidden* because they can only be inferred through observations. Some paths are controllable, in that they can be realised through action. Therefore, certain observations depend upon action or policies, which requires the generative model to entertain expectations about the consequences of action. These expectations are optimised by minimising *variational free energy* while the prior probability of a policy depends upon its *expected free energy* (Parr et al., 2022).

Figure 1 provides an overview of the generative models typically used for sense and decision-making under uncertainty; namely, a partially observed Markov decision process (POMDP). Outcomes at any time depend upon hidden *states*, while transitions among hidden states depend upon *paths*. In this model, paths are random variables that depend upon action. This POMDP is specified by a set of tensors. The first set **A**, maps from hidden states to outcome modalities; for example, exteroceptive (e.g., visual) or proprioceptive (e.g., eye position) *modalities*. These parameters encode the likelihood of an outcome given hidden states. The second set **B** encodes transitions among the hidden states of a *factor*, under a particular path. Factors correspond to different kinds of states; e.g., the location versus the class of an object. The remaining tensors encode prior beliefs about paths **C**, and initial conditions **D** and **E**, sometimes referred to as hidden *causes*. The tensors are generally parameterised as Dirichlet distributions, whose sufficient statistics are *Dirichlet counts*, which count the number of times a particular combination of states and outcomes has been inferred. We will focus on learning the likelihood model, encoded by Dirichlet counts, **a**.

Artificial reasoning

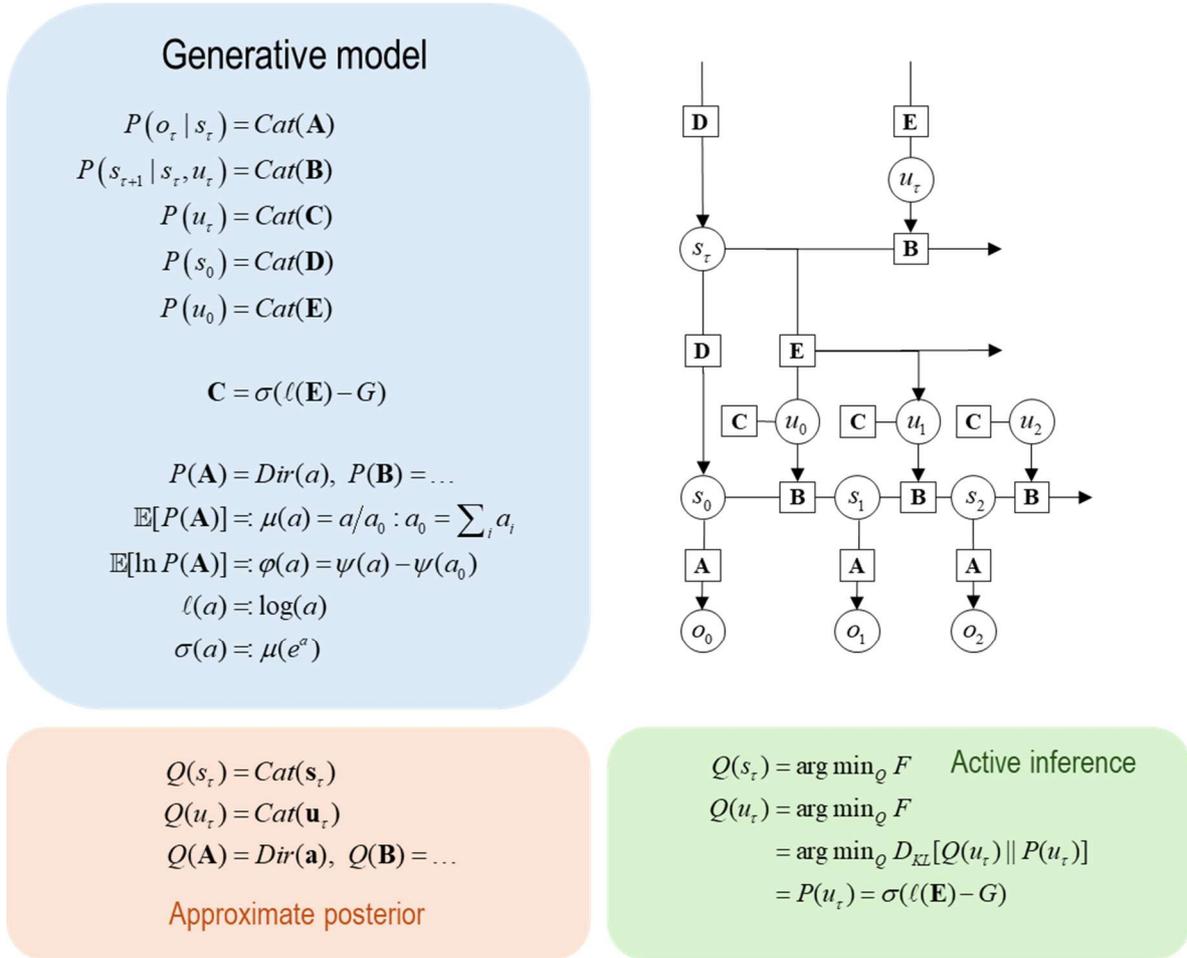

**Figure 1**: **Generative models.** A generative model specifies the joint probability of observable consequences and their hidden causes. Usually, the model is expressed in terms of a *likelihood* (the probability of consequences given their causes) and *priors* (over causes). When a prior depends upon a random variable it is called an *empirical prior*. Here, the likelihood is specified by a tensor **A**, encoding the probability of an outcome under every combination of *states* (*s*). Priors over transitions among hidden states, **B** depend upon *paths* (*u*), whose transition probabilities are encoded in **C**, where **D** and **E** specify the prior over initial states and paths, respectively. Certain paths are more probable *a priori* if they minimise their expected free energy (*G*). For clarity, we have assumed that there is a single factor and path in this model. The blue panel provides the functional form of the generative model in terms of categorical (*Cat*) distributions that are parameterised as Dirichlet (*Dir*) distributions. The lower equalities list the various operators required for variational message passing in Figure 2. These functions operate on each column of their tensor arguments. $\psi(\cdot)$ is the digamma function. The right panel depicts the generative model as a probabilistic graphical model and the lower panels list the functional form of approximate posteriors (*Q*) and their optimisation.

The generative model in Figure 1 means that outcomes are generated by selecting a policy from a softmax function of expected free energy. Sequences of hidden states are then generated using the probability transitions specified by the selected combination of paths (i.e., policy). Finally, hidden states generate outcomes in one or more modalities. Model inversion updates the sufficient statistics $(\mathbf{s}, \mathbf{u}, \mathbf{a}, \ldots)$ of posterior beliefs $Q(s, u, \mathbf{A}, \ldots) = Q(s)Q(u)Q(\mathbf{A}), \ldots$ that are factorised over hidden states, paths and parameters. This mean field factorisation effectively partitions belief updating into *inference*, *planning* and *learning*.

## Variational and expected free energy

In variational (a.k.a., approximate) Bayesian inference, model inversion entails the minimisation of variational free energy with respect to the sufficient statistics of the approximate posterior. For clarity, we will deal with a single factor, such that the policy (i.e., combination of paths) becomes the path. Omitting dependencies on initial states and certain parameters, for model $m$ we have:

$$Q(s_\tau, u_\tau, \mathbf{A}) = Q(s_\tau)Q(u_\tau)Q(\mathbf{A}) = \arg\min_Q F$$

$$\begin{aligned}
F &= \mathbb{E}_Q[\ln \underbrace{Q(s_\tau, u_\tau, \mathbf{A})}_{\text{Posterior}} - \ln \underbrace{P(o_\tau \mid s_\tau, \mathbf{A})}_{\text{Likelihood}} - \ln \underbrace{P(s_\tau \mid s_{\tau-1}, u_\tau)P(u_\tau)P(\mathbf{A})}_{\text{Prior}}] \\
&= \underbrace{\mathbb{E}_Q[D_{KL}[Q(s_\tau, u_\tau, \mathbf{A}) \,\|\, P(s_\tau, u_\tau, \mathbf{A} \mid s_{\tau-1}, o_\tau)]]}_{\text{Divergence}} - \underbrace{\ln P(o_\tau \mid m)}_{\text{Log evidence}} \\
&= \underbrace{\mathbb{E}_Q[D_{KL}[Q(s_\tau, u_\tau, \mathbf{A}) \,\|\, P(s_\tau, u_\tau, \mathbf{A} \mid s_{\tau-1})]]}_{\text{Complexity}} - \underbrace{\mathbb{E}_Q[\ln P(o_\tau \mid s_\tau)]}_{\text{Accuracy}}
\end{aligned} \quad (1)$$

Here, the expectation is under $Q = Q(s_{\tau-1})Q(s_\tau, u_\tau, \mathbf{A})$. Because the (KL) divergences cannot be less than zero, the penultimate equality means that free energy is minimised when the approximate posterior is the true posterior. At this point, the free energy becomes the negative log evidence for the generative model (Beal, 2003). This means minimising free energy is equivalent to maximising model evidence. Planning emerges under active inference by placing priors over paths to minimise expected free energy (Friston et al., 2015). In the simple case of one-step-ahead policies (and limiting our focus to the likelihood parameters), we have:

$$\begin{aligned}
G(u) &= \mathbb{E}_Q[\ln Q(s_{\tau+1}, \mathbf{A} \mid u) - \ln Q(s_{\tau+1}, \mathbf{A} \mid o_{\tau+1}, u) - \ln P(o_{\tau+1} \mid c)] \\
&= -\underbrace{\mathbb{E}_Q[\ln Q(\mathbf{A} \mid s_{\tau+1}, o_{\tau+1}, u) - \ln Q(\mathbf{A} \mid s_{\tau+1}, u)]}_{\text{Expected information gain (Parameters)}} - \\
&\quad \underbrace{\mathbb{E}_Q[\ln Q(s_{\tau+1} \mid o_{\tau+1}, u) - \ln Q(s_{\tau+1} \mid u)]}_{\text{Expected information gain (States)}} - \underbrace{\mathbb{E}_Q[\ln P(o_{\tau+1} \mid c)]}_{\text{Expected cost}} \\
&= -\underbrace{\mathbb{E}_Q[D_{KL}[Q(\mathbf{A} \mid s_{\tau+1}, o_{\tau+1}, u) \,\|\, Q(\mathbf{A} \mid s_{\tau+1}, u)]]}_{\text{Novelty}} + \\
&\quad \underbrace{D_{KL}[Q(o_{\tau+1} \mid u) \,\|\, P(o_{\tau+1} \mid c)]}_{\text{Risk}} - \underbrace{\mathbb{E}_Q[\ln Q(o_{\tau+1} \mid s_{\tau+1}, u)]}_{\text{Ambiguity}}
\end{aligned} \quad (2)$$

Artificial reasoning

Here, $Q =: Q(o_{\tau+1}, s_{\tau+1}, \mathbf{A} | u) = P(o_{\tau+1}, s_{\tau+1}, \mathbf{A} | u, o_0, \ldots, o_\tau) = P(o_{\tau+1} | s_{\tau+1}, \mathbf{A}) Q(s_{\tau+1}, \mathbf{A} | u)$ is the posterior predictive distribution over parameters, hidden states and outcomes at the next time step, under a particular path. Note that the expectation is over *observations in the future*, hence, *expected* free energy. This means that preferred outcomes— parameterised by *c*—are prior beliefs, which constrain the implicit planning as inference: c.f., (Attias, 2003; Botvinick and Toussaint, 2012; Van Dijk and Polani, 2013). Planning over longer time horizons simply involves accumulating the path integral of expected free energy over successive actions—or combinations of actions; i.e., policies—as described in (Friston et al., 2021).

Expected free energy augments expected information gain with costs or constraints. These constraints reflect the fact that we are dealing with systems with characteristic outcomes that are scored by their cost $\mathcal{C}(o) = -\ln P(o|c)$ or inverse value. Statistically speaking, expected free energy underwrites Bayes optimal behaviour, in the dual sense of Bayesian decision theory (Berger, 2011) and optimal experimental design (Lindley, 1956). Its functional form inherits from the path integral formulation of the free energy principle: please see (Friston et al., 2023) for details. One can also express a prior over parameters in terms of an expected free energy where, marginalising over paths:

$$P(\mathbf{A}) = \sigma(-G(\mathbf{A}))$$

$$G(\mathbf{A}) = \mathbb{E}_Q[\ln P(s|\mathbf{A}) - \ln P(s|o, \mathbf{A}) - \ln P(o|c)]$$

$$= -\underbrace{\mathbb{E}_Q[\ln P(s|o,\mathbf{A}) - \ln P(s|\mathbf{A})]}_{\text{Expected information gain}} - \underbrace{\mathbb{E}_{Q_a}[\ln P(o|c)]}_{\text{Expected cost}}$$

$$= -\underbrace{\mathbb{E}_Q[D_{KL}[P(o,s|\mathbf{A}) \| P(o|\mathbf{A})P(s|\mathbf{A})]]}_{\text{Mutual information}} - \underbrace{\mathbb{E}_{Q_a}[\ln P(o|c)]}_{\text{Expected cost}} \quad (3)$$

Here, $Q =: P(o|s,a)P(s|a) = P(o,s|a)$ is the joint distribution over outcomes and hidden states, encoded by Dirichlet parameters. Note that the Dirichlet parameters encode the mutual information, because they encode the joint distribution over outcomes and hidden states: when normalising each column of the *a* tensor, we recover the likelihood distribution. However, we could normalise over all entries of the tensor to recover a joint distribution. The prior in (3) can be read as implementing a constrained principle of maximum mutual information or minimum redundancy (Ay et al., 2008; Barlow, 1961; Linsker, 1990; Olshausen and Field, 1996). In machine learning, this kind of prior underwrites disentanglement (Higgins et al., 2021; Sanchez et al., 2019), and generally leads to sparse representations (Gros, 2009; Olshausen and Field, 1996; Sakthivadivel, 2022; Tipping, 2001). We will call on this prior later when specifying priors over models.

## Active inference

In variational inference, sufficient statistics—encoding posterior expectations—are updated to minimise variational free energy. Figure 2 illustrates these updates in the form of variational message passing (Dauwels, 2007; Winn and Bishop, 2005). For example, expectations about hidden states are a softmax function of messages that are linear combinations of other expectations and observations.

$$\begin{aligned} \mathbf{s}_\tau^f &= \sigma(\mu_{\uparrow \mathbf{A}}^f + \mu_{\rightarrow B}^f + \mu_{\leftarrow B}^f + \ldots) \\ \mu_{\uparrow \mathbf{A}}^f &= \sum_{g \in ch(f)} \mu_{\uparrow \mathbf{A}}^{g,f} \\ \mu_{\uparrow \mathbf{A}}^{g,f} &= \mathbf{o}_\tau^g \odot \varphi(\mathbf{a}^g) \odot_{i \in pa(g) \setminus f} \mathbf{s}_\tau^i \end{aligned} \qquad (4)$$

Here, the ascending messages from the likelihood factor are a linear mixture[1] of expected states and observations, weighted by digamma functions of the Dirichlet counts that parameterise the likelihood model (c.f., connection weights). The expressions in Figure 2 are effectively the fixed points (i.e., minima) of variational free energy. This means that message passing corresponds to a fixed-point iteration scheme (Beal, 2003; Dauwels, 2007; Winn and Bishop, 2005).[2]

---

[1] The $\odot$ notation implies a sum product operator; i.e., the dot or inner product that sums over one dimension of an array or tensor. These operators are applied to a vector $\mathbf{a}$ and tensor $\mathbf{A}$ where, $\mathbf{a} \odot \mathbf{A}$ implies the sum of products is taken over the leading dimension, while $\mathbf{A} \odot \mathbf{a}$ implies the sum is taken over a trailing dimension. For example, $\mathbf{1} \odot \mathbf{A}$ is the sum over rows and $\mathbf{A} \odot \mathbf{1} = \mathbf{A}_0$ is the sum over columns, where $\mathbf{1}$ is a vector of ones and $\mathbf{A}$ is a matrix. If $\mathbf{A}$ is a matrix then $\mathbf{a} \odot \mathbf{A} = \mathbf{a}^T \cdot \mathbf{A}$. Finally, $\mathbf{A}_{\bullet,i}$ and $\mathbf{A}_{i,\bullet}$ refer to the $i$-th column and row of $\mathbf{A}$, respectively. This notation replaces the Einstein summation notation to avoid visual clutter.

[2] Figure 2 and Equation (4) provide forward and backward messages in time. In practice—and in the examples below—it is sometimes simpler to omit backward messages; in which case the scheme becomes a Bayesian filter—as opposed to smoother. One can also replace the variational messages with those from sum-product belief-propagation schemes to achieve exact marginal posteriors.

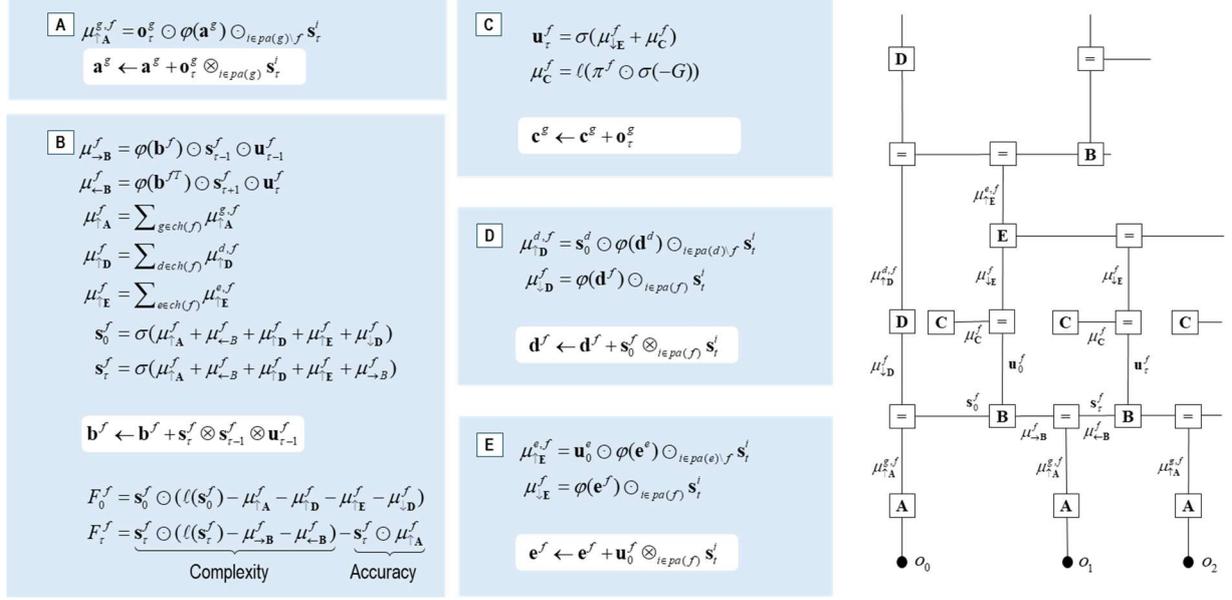

**Figure 2**: **Belief updating and variational message passing**: the right panel presents the generative model as a Forney factor graph, where the nodes (square boxes) correspond to the factors of the generative model (labelled with the associated tensors). The edges connect factors that share dependencies on random variables. The leaves (filled circles) correspond to known variables, such as observations (*o*). This representation is useful because it scaffolds the message passing—over the edges of the factor graph—that underwrite inference and planning. The functional forms of these messages are shown in the left-hand panels. For example, the expected path—in the first equality of panel **C**—is a softmax function of two messages. The first is a descending message $\mu^{f}_{\downarrow E}$ from **E** that inherits from expectations about hidden states at the level above. The second is the log-likelihood of the path based upon expected free energy, *G*. This message depends upon Dirichlet counts scoring preferred outcomes—i.e., prior constraints on modality *g*—encoded in $\mathbf{c}^{g}$ : see Equation (2). The matrix $\pi$ encodes the probability over paths, under each *policy*. Subscripts pertain to time, while superscripts denote distinct factors *f*, and outcome modalities *g*. Tensors and matrices are denoted by uppercase bold, while posterior expectations are in lowercase bold. The ⊙ notation denotes a generalised inner (i.e., dot) product or tensor contraction, while ⊗ denotes an outer product. The updates in the lighter panels correspond to learning; i.e., updating Bayesian beliefs about parameters. $ch(\cdot)$ and $pa(\cdot)$ return the children and parents of latent variables.

Belief updating under active inference is sample efficient by design, in virtue of acting to maximise information gain—and minimise ambiguity—where expected free energy can be expressed in terms of sufficient statistics as follows:

Artificial reasoning

$$G_h = \sum_g \underbrace{\mathbf{o}^{g,h}_{\tau+1} \odot \ell(\mathbf{o}^{g,h}_{\tau+1}) - \mathbf{o}^{g,h}_{\tau+1} \odot \varphi(\mathbf{c}^g)}_{\text{Risk}} \underbrace{- \mathbf{H}^g \odot \mathbf{s}^h_{\tau+1}}_{\text{Ambiguity}} \underbrace{- \mathbf{o}^{g,h}_{\tau+1} \cdot \mathbf{W}^g \cdot \mathbf{s}^h_{\tau+1}}_{\text{Novelty}} \quad (5)$$

$$\mathbf{o}^{g,h}_{\tau+1} = \mu(\mathbf{a}^g) \odot \mathbf{s}^h_{\tau+1}$$

$$\mathbf{s}^h_{\tau+1} = \mu(\mathbf{b} \odot \pi_h) \odot \mathbf{s}_\tau$$

$$\mathbf{H}^g = \mathbf{1} \odot (\mu(\mathbf{a}^g) \times \varphi(\mathbf{a}^g))$$

$$\mathbf{W}^g_{ij} = \frac{1}{\mathbf{a}^g_{ij}} - \frac{1}{\mathbf{a}^g_{0j}} + \ln \frac{\mathbf{a}^g_{0j}}{\mathbf{a}^g_{ij}} + \psi(\mathbf{a}^g_{ij}) - \psi(\mathbf{a}^g_{0j}) \approx \frac{1}{2\mathbf{a}^g_{ij}} - \frac{1}{2\mathbf{a}^g_{0j}}$$

This expresses expected free energy in terms of risk, ambiguity and novelty for policy $h$; namely, the $h$-th combination of actions over factors.

## Active learning

In the setting of discrete models, learning corresponds to updating model parameters by accumulating Dirichlet counts based upon posterior expectations. For example, for the likelihood tensors we have:

$$\mathbf{a}^g \leftarrow \mathbf{a}^g + \mathbf{o}^g_\tau \otimes_{i \in pa(g)} \mathbf{s}^i_\tau \quad (6)$$

Effectively, the likelihood parameters count the number of co-occurrences of outcomes and their latent causes. The expected information gain over parameters in (2) and (5) ensures that outcomes are sampled to minimise uncertainty about the likelihood mapping. Intuitively, this implies seeking outcomes that inform columns of the likelihood tensor that have small Dirichlet counts. Conversely, when combinations of latent states and outcomes have been seen many times, there is less uncertainty to resolve and the outcomes offer less epistemic affordance or expected information gain.

## Active selection

In contrast to learning—that optimises *posteriors* over parameters—Bayesian model selection or structure learning (Tenenbaum et al., 2011; Tervo et al., 2016; Tomasello, 2016) can be framed as optimising the *priors* over model parameters. On this view, model selection can be implemented efficiently using Bayesian model reduction. Bayesian model reduction is a generalisation of ubiquitous procedures in statistics, such as the Savage–Dickey method (Savage, 1954). By applying Bayes rules to full and reduced models, it is straightforward to show that the change in log marginal likelihood can be expressed in terms of posterior Dirichlet counts $\mathbf{a}$, prior counts $a$ and the prior counts that define a reduced model $a_m =: a \mid m$. Using $B$ to denote the beta function, we have (Friston et al., 2018):

Artificial reasoning

$$\begin{aligned}\mathcal{B}(\mathbf{a}, a_m, a) &= \ln P(o \mid a) - \ln P(o \mid a_m) \\ &= \ln \mathrm{B}(\mathbf{a})\mathrm{B}(a_m) - \ln \mathrm{B}(a)\mathrm{B}(\mathbf{a}_m) \\ \mathbf{a}_m &= \mathbf{a} + a_m - a\end{aligned} \qquad (7)$$

Here, $\mathbf{a}_m$ corresponds to the posterior that would obtain under the reduced model.

So far, we have not considered the priors over parameters. By appeal to (3), we can assume a set of models $m \in \mathcal{M}$, all of which entail a high mutual information (between latent states and observations), under suitable constraints. In this setting, the model $m$ becomes a random variable, over which we can assume a uniform prior. This leads to a prior over parameters that can be read as a Bayesian model average:

$$\begin{aligned}a &= \mathbb{E}_{P(m)}[a_m] \\ P(\mathbf{A}) &= Dir(a) \\ P(m) &= 1/|\mathcal{M}|\end{aligned} \qquad (8)$$

This specification of priors admits a simple evaluation of posteriors over models in terms of the Bayesian model reduction of Dirichlet parameters that have accumulated outcomes in the past. Applying Bayes rule and dropping terms that do not depend upon $m$, we have:

$$\begin{aligned}\ln P(m \mid o) &= \ln P(o \mid m) + \ln P(m) - \ln P(o) \\ &= \ln P(o \mid a) - \mathcal{B}(\mathbf{a}, a_m, a) - \ln |\mathcal{M}| - \ln P(o) \\ &\Rightarrow \\ P(m \mid o) &= \sigma(-\mathcal{B}) \\ \mathcal{B}_m &= \mathcal{B}(\mathbf{a}, a_m, a) = \ln P(o \mid a) - \ln P(o \mid m)\end{aligned} \qquad (9)$$

In summary, if we start with a set of models or hypothesis space $\mathcal{M}$, of equally plausible models (here, likelihood mappings), we have a principled way of specifying the prior over model parameters in terms of a Bayesian model average. One can then accumulate evidence for each of the models until one is sufficiently confident that one model supervenes over the remainder. One then selects the 'last man standing' for subsequent inference and learning; i.e., applies Occam's razor. This constitutes Bayesian model selection by reduction. See (Kiebel and Friston, 2011) for a biomimetic example of this form of self-organisation, applied to dendritic neuronal processes and the sampling of presynaptic inputs. So, can we optimise evidence accumulation for subsequent model selection?

Using the expression for posteriors models in (9), one can generalise expected free energy to include the expected information gain *over models*. This rests upon equipping an agent with a repertoire of models or hypotheses, so that it can actively search for disambiguating evidence and identify the most plausible model through an application of Occam's razor. The expected information gain is relatively straightforward to evaluate, given the evidence accumulated by the Dirichlet parameterisation of the full or parent model in (8). The expected information gain is the expected KL divergence between the posterior probability of a model with, and without, the outcomes under a particular policy.

$$
\begin{aligned}
G(u) = &-\underbrace{\mathbb{E}_Q[\ln Q(m \mid \mathbf{A}, s_{\tau+1}, o_{\tau+1}, u) - \ln Q(m \mid \mathbf{A}, s_{\tau+1}, u)]}_{\text{Expected information gain (Models)}} \\
&-\underbrace{\mathbb{E}_Q[\ln Q(\mathbf{A} \mid s_{\tau+1}, o_{\tau+1}, u) - \ln Q(\mathbf{A} \mid s_{\tau+1}, u)]}_{\text{Expected information gain (Parameters)}} \\
&-\underbrace{\mathbb{E}_Q[\ln Q(s_{\tau+1} \mid o_{\tau+1}, u) - \ln Q(s_{\tau+1} \mid u)]}_{\text{Expected information gain (States)}} \\
&-\underbrace{\mathbb{E}_Q[\ln P(o_{\tau+1} \mid c)]}_{\text{Expected cost}}
\end{aligned}
\qquad (10)
$$

The requisite predictive posteriors for the expected information gain over models follow from equations (5), (6) and (9):

$$
\begin{aligned}
Q(m \mid \mathbf{A}, s_{\tau+1}, o_{\tau+1}, u) &= \sigma(-\mathcal{B}) : \mathcal{B}_m = \mathcal{B}(\mathbf{a}^g + \Delta^g, a_m^g, a^g) \\
Q(m \mid \mathbf{A}, s_{\tau+1}, u) &= \sigma(-\mathcal{B}) : \mathcal{B}_m = \mathcal{B}(\mathbf{a}^g, a_m^g, a^g) \\
\Delta^g &= \mathbf{o}_{\tau+1}^{g,u} \otimes_{i \in pa(g)} \mathbf{s}_{\tau+1}^{u}
\end{aligned}
\qquad (11)
$$

Note that *Q* here is a predictive posterior, as opposed to an approximate posterior over models. In other words, it is the predictive posterior under the model average in (8). This speaks to the distinction between the current *post-hoc* application of Bayesian model reduction (Friston and Penny, 2011) and inferring the posterior over models through continuous structure learning. Please see the appendix for a brief discussion.

For clarity, we have assumed only the likelihood mapping to modality *g* is unknown and that there is only one factor with a controllable path (i.e., $h = u$). When multiple mappings are unknown, one simply adds the accompanying expected information gains to expected free energy. Note that this expected information gain does not lend itself to evaluation over sequences of actions, unless the parameters are updated over successive time points in the future. Therefore, we will restrict the use of expected information gain over models to the next action—or combination thereof—and invoke planning over longer time horizons after the best model has been selected.

Including the expected information gain over models affords the evidence or data that best disambiguates hypotheses or models. It does not, in and of itself, commit to using a new model. To replace the full model with a reduced model requires an application of Occam's razor when, and only

# Artificial reasoning

when, the agent is sufficiently confident that a reduced model is the best explanation for the evidence accumulated thus far. One can operationalise this application of Occam's razor in terms of a log Bayes factor (i.e., log odds ratio) scoring the probability that the best model is more likely than any other model:

$$\mathcal{R} = \ln \frac{P(m^*|o)}{1 - P(m^*|o)} \tag{12}$$
$$m^* = \arg\max_m P(m|o)$$

When Occam's razor surpasses a threshold, the agent replaces its posterior over parameters with the Bayesian model average based upon its posterior over models. If we use a threshold of 16 natural units, the Bayesian model average becomes Bayesian model selection. This follows because the winning model is at least exp(16) times more likely than alternative models:

$$\mathbf{a} \leftarrow a_{m^*} = \mathbb{E}_{P(m|o)}[a_m] \Rightarrow Q(\mathbf{A}) = Dir(a_{m^*}) \tag{13}$$

In practice, one can multiply the Dirichlet counts in the selected model by some arbitrarily high number (e.g., 512), to suppress novelty seeking. Conversely, one can render the agent impressionable, and use the reduced model directly (whose Dirichlet counts usually sum to one over columns).

The ensuing active model reduction or reasoning recapitulates how a scientist would proceed when resolving uncertainty about her alternative hypotheses. She will acquire a sufficient amount of data—under an unbiased (flat) prior over her hypotheses—to inform her predictive posterior. After acquiring the data, she will commit to the best hypothesis, if the evidence in favour of this hypothesis is sufficiently large, relative to others. This analogy emphasises the separation of temporal scales that accompanies active inference, learning and selection. In short, inference is a fast process of belief updating about time-dependent latent states. Learning is a slow process of belief updating about the parameters of a generative model, and selection is a *post-hoc* process that is enacted after sufficient data have been accumulated.

In this *post-hoc* scheme, evidence is accumulated by the Dirichlet parameters that are learned; enabling Bayesian model selection based purely upon what has observed thus far. One could imagine invoking hyperpriors about the number of data points required before selecting a model. Indeed, scientists are often required to do power calculations to estimate the number of observations that should be collected before analysis. For simplicity, we will just use the log odds ratio in equation (12) , to select the best model—i.e., apply Occam's razor—when it surpasses a threshold. Clearly, this does not guarantee that the selected model is the best model, in the sense that the agent could jump to premature conclusions. However, these should be relatively rare events with a sufficiently high threshold.

Artificial reasoning

An outstanding issue is the elaboration of alternative models; i.e., the specification of the model space. The requisite specification of $\mathcal{M}$ effectively implements hyperpriors over hypothetical rules, symmetries, invariances and the like. The problem chosen to illustrate the application of the above scheme speaks to a fairly straightforward procedure for automatically specifying the model space. We will return to this specification after describing the problem in more detail.

## THE THREE-BALL PARADIGM

The three-ball problem was originally used to introduce active learning and model selection as a mathematical homologue of insight and 'aha moments' that evince (artificial) curiosity (Friston et al., 2017). Here, we use a nuanced version of this paradigm to demonstrate the active sampling of data that renders Bayesian model reduction as efficient as possible, through supplementing expected free energy with expected information gain over models.

In brief, an agent is presented with three coloured balls and invited to choose a colour. If the agent makes a choice (e.g., through a button press), then it receives feedback that it was correct or incorrect. The agents prior preferences, $c$, are set such that the relative log probabilities of no feedback, correct and incorrect are 0, 2 and -6 natural units, respectively. In other words, it prefers to receive correct feedback and finds an incorrect outcome highly surprising and therefore aversive. The rule that the agent has to discover—to secure preferred outcomes—is context sensitive: if the centre ball is green, then the correct colour is green. However, if the centre ball is red the correct colour is on the left, while if the centre ball is blue, the correct colour is on the right. This rule is installed in the likelihood mapping between latent states encoding the colours of the three balls—and the colour of choice—to the feedback modality. Therefore, the agent must learn and select the likelihood tensor that encodes the contingencies constituting the rule. So, how might the agent discover this rule?

If the agent could see all three balls at the same time, then it would be a relatively straightforward matter to *learn* the rule using reinforcement learning; e.g., (Silver et al., 2021). For example, there are only 3×3×3×4 = 108 combinations of ball colours and responses (i.e., three colour choices and a 'no response' option). This means that after 500 training trials the probability of seeing at least one instance of every combination is at least 99% $=1 – (1 – 1/108)^{500}$ and the likelihood mapping could be learned almost certainly. However, problems like the three-ball paradigm preclude a learnable state-action policy because the agent can only see one ball at a time, depending upon where it is looking. This converts a Markov decision problem into a partially observed problem, where the agent not only has to choose the correct colour but also choose where to look. This ubiquitous kind of problem calls for sequential policy optimisation. Intuitively, knowing that there is a red ball on the left does not specify the correct response because the centre ball could be green or blue. This means that the optimal policy is to first resolve uncertainty about the context (by looking at the centre ball) and respond immediately (if it is green) or look to the right or left (if it is blue or red). And then choose the observed colour. The ensuing problem can now be specified with the following generative model.

The latent states of this model are fivefold. There are three sorts of states (i.e., factors) encoding the colour of the ball at each *location* (we will refer to these as *location* factors). The remaining (controllable) factors are *where* the agent is looking (*start*, *left*, *centre*, or *right*) and the agent's *choice* (*no choice*, *red*, *blue* or *green*). These hidden states are sufficient to generate two outcome modalities; namely, a single *visual* modality (seeing *nothing*, a *red*, *green* or *blue* ball) and a *feedback* modality (*no feedback*, *correct*, *incorrect*). The rule is encoded in the likelihood tensor $\mathbf{A}^g \in \mathbb{R}^{3 \times 3 \times 3 \times 4 \times 4}$ mapping from hidden states to the *feedback* modality. The transition priors for this model are very simple. There are no transitions among the colours of the balls at each *location* within a given trial, so the transition tensors are identity matrices $\mathbf{B}^f = I \in \mathbb{R}^{3 \times 3}$. The controllable or action-dependent factors depend upon action, $\mathbf{B}^f \in \mathbb{R}^{4 \times 4 \times 4}$; e.g., looking to the *left* moves the location state to *left*, and so on. Similarly for the *choice* factor. On each trial the colours of the balls are selected at random from $\mathbf{D}^f \in \mathbb{R}^3$ with the agent looking at the *start* location. Policies are selected on the basis of expected free energy and prior preferences over the feedback modality are specified by $\mathbf{c}^g = \sigma([0, 2, -6]) \in \mathbb{R}^3$. The agent is allowed to forage and choose for 6 time points, after which a new trial starts, with randomly selected ball colours.

Figure 4 illustrates the behaviour of an agent equipped with the above generative model and the correct rule. This figure reports belief updating and choice behaviour in terms of the posterior densities over latent states and policies, in image format. In this example, the colour of the centre ball was red, meaning the correct choice was the colour of the red ball on the left. With informative preferences and active inference—looking two moves into the future—the agent first resolves uncertainty about the colour of the informative ball by looking from the *start* location to the *centre* ball. Having resolved its uncertainty about the context in play, it looks at the *left* ball to infer its colour. Following this, it then *chooses* the red option for the remaining moves, thereby maximising the number of preferred (i.e., *correct*) outcomes. However, while continuing to *choose* red, it resolves further uncertainty about the uninformative ball on the right. In other words, it continues visual *exploration* at the same time as *exploiting* its posterior beliefs about the correct response. The question we now address is: what would happen if the agent did not know the rule? And what kind of prior should it use in the absence of this knowledge?

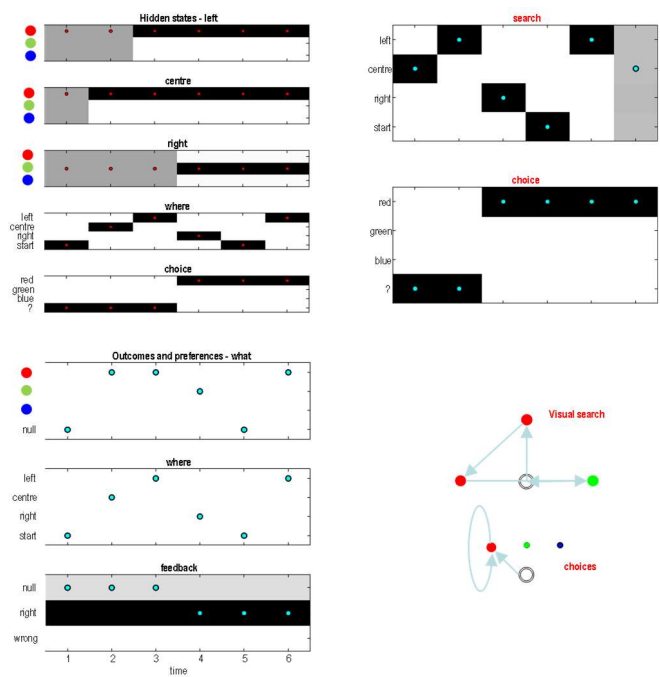

**Figure 3: choice behaviour**. This figure reports belief updating for a single trial under the three-ball paradigm. The panels on the upper left report posterior beliefs in image format over five latent state factors and six time points. Two of these latent factors are controllable and are equipped with posteriors over paths, with four levels each. These posteriors are shown on the upper right panel. The lower left panel reports the outcomes solicited by the agent. Of these outcomes, the third (*feedback*) outcome has informative prior preferences, such that the agent prefers to be *correct* or right and finds being *incorrect* or wrong highly surprising. Darker colours indicate higher probability and the dots represent the actual states and outcomes. The lower right panel illustrates the response in terms of where the agent was looking (Visual search) and choice behaviour (Choices). In this example, the agent first looks to the centre location and observes the colour is red. It then knows that the informative (*criterion*) location is on the left. Looking to the left, it observes a red ball and subsequently *chooses* red. The agent then looks to the right to resolve the remaining uncertainty about the colour of the right ball. Once the correct colour has been inferred, the agent continues to choose this colour, to maximise preferred feedback.

For simplicity, we will assume that the generative model has been learned by the agent, with the exception of the likelihood tensor mapping from latent states to the *feedback* modality. It is this likelihood mapping that contains the rule the agent has to discover. So, what alternative likelihood mappings could the agent entertain?

When specifying (or inheriting) priors over the parameters of a generative model, one is implicitly specifying a set of reduced models that can be derived by removing parameters from the full prior. Alternatively, one could argue that the initial priors are, effectively, a mixture—or Bayesian model average—over reduced models that all satisfy some constraints. But what constraints? One can now turn to the hyperprior in (3) and note that a necessary condition for a maximum mutual information is that each column of the likelihood tensor should have a unique nonzero entry[3]. This also ensures maximum sparsity, in the sense that the number of zero entries equals its supremum.

Sparsity is perhaps the simplest constraint, where all models feature a unique mapping from every combination of latent states to some response; e.g., *correct* or *incorrect*. This could be a minimal definition of a rule. For example, if we consider the likelihood tensor as a joint distribution over latent states and outcomes, one can specify priors by conditioning on a *correct* outcome and consider the structure of the corresponding subtensor: e.g., $\mathbf{A}^g_{correct} \in \mathbb{R}^{1 \times 3 \times 3 \times 4 \times 4}$. In the current example, there are $3 \times 3 \times 3 \times 4 \times 4 = 432$ combinations of latent states that can be assigned a concentration parameter of one or zero, depending upon whether any given combination leads to a *correct* or *incorrect* outcome, respectively. This would lead to $2^{432}$ models or rules. Clearly, this would be a weak kind of a rule, in the sense a rule should compress explanations for observations into a small set, in the spirit of minimising algorithmic complexity or description length: e.g., (Hinton and Zemel, 1993; Ruffini, 2017; Schmidhuber, 2010; Wallace and Dowe, 1999). In short, sparsity is a necessary but not sufficient condition for maximising mutual information. To maximise the mutual information, it is necessary for some combinations of hidden states to generate a correct outcome and others not. Beyond this, the mutual information constraint tells us little about plausible rules.

However, the latent states have a factorial structure that distinguishes between states that can be controlled or chosen and those that cannot. We can refer to these as *choice* and *criterion* states, respectively. One could consider a rule to implement the constraint that, for any combination of *criterion* states, there is a unique *choice* state that generates a true outcome. This guarantees maximum mutual information. For example, there are $3 \times 3 \times 3 = 27$ combinations of *criterion* (i.e., *location*) states, each of which can be associated with one of $4 \times 4 = 16$ *choice* states. This reduces the number of rules to $16^{27} = 2^{31}$, which is still greater than Avogadro's number. So, can we go further and impose more symmetries to define a model space?

One can now consider symmetries among factors and hypothesise an isomorphic mapping between a *choice* and *criterion* state if, and only if, they share the same support or domain. For example, here, the three location factors (*left*, *centre* and *right*) share annotated states with the choice factor; namely, (*red*, *blue* and *green*). This isomorphism may or may not depend upon the *context*; namely, a combination of criterion states. In other words, for every feedback outcome, there is a slice of the likelihood tensor (chosen such that one dimension is taken from the *choice* factors, and the other from the *criterion* factors) that has the form of an identity mapping. This, in the current example, leads to 3×3×3 = 27 contexts that could specify one of three criterion states that would support an isomorphism with a choice state. This constraint gives $3^{27}$ models, which is still too large. So, can we turn to weaker—i.e., sparsity—constraints?

---

[3] One can demonstrate this explicitly by noting that mutual information between states and observations (given likelihood parameters) can be decomposed into the prior entropy of states minus the conditional entropy of states given observations. Only the latter depends upon the likelihood parameters. This means the maximum mutual information is obtained when the conditional entropy of states, for any plausible observation, is zero—i.e., when there is a single plausible state for each observation. With multiplication by the prior vector for states and renormalisation, one can attain the sparsity constraints for the likelihood matrix (or tensor).

A weaker symmetry or isomorphism pertains to the size of hidden state factors. Here, there are several *context* states that could uniquely specify a *criterion* state, in virtue of the fact that the size of each *context* state equals the number of criterion states. Recall the *context* states are the set of *criterion* states and that, in our setup, there are 3 *criterion* state factors comprising the *context* with 3 possible values each of the *context* states might take (red, green, or blue). One could now hypothesise that a context state specifies a unique criterion state, thereby satisfying a sparsity constraint. In the current example, there are three possible criterion states (left, centre, or right), each of which could be specified by one of three context states (red, green, or blue) in $3^3$ ways (e.g., left red → left, left red → centre, …, centre red → left, centre red → centre, …, right blue → right). This leads to $3 \times 3^3 = 81$ possible rules[4] (3 possible values the indicated *criterion* state factor might take by the $3^3$ ways of specifying a mapping from *context* to criterion). These symmetry constraints now provide a plausible hypothesis space, $\mathcal{M}$ that (i) jointly specifies the full prior according to (8), (ii) affords the opportunity for active reduction or reasoning via expected information gain in (10) and (iii) subsequent model selection according to (12).

Generally speaking, rules or likelihood models can be specified automatically—in terms of Dirichlet tensors—by identifying a (controlled) *choice* factor that shares more than one annotated state with (noncontrolled) *criterion* factors. One then identifies candidate *context* factors whose levels can uniquely specify the criterion factor. Finally, for every $n \times m^m$ combination of *n* context and *m* criterion factors, a putative likelihood mapping is generated by assigning a Dirichlet account to a *correct* outcome, whenever the *choice* matches the *criterion* state specified by the *context* state. Algorithmically, this can be expressed as follows, where, $s_i^f$ denotes the level of factor *f* under the *i*-th combination of states:

For *n* context states and $m^m$ specifications of *m* criterion states

    For every combination of states, *i*

---

[4] Some of these rules will be degenerate, in the sense there are only 79 unique rules.

Artificial reasoning

$$criterion = s_i^{context}$$

$$a_{correct,s_i^1,s_i^2,\ldots}^{feedback} = \begin{cases} 1 & s_i^{choice} = s_i^{criterion} \\ 0 & s_i^{choice} \neq s_i^{criterion} \end{cases} \qquad (14)$$

$$a_{incorrect,s_i^1,s_i^2,\ldots}^{feedback} = \begin{cases} 0 & s_i^{choice} = s_i^{criterion} \\ 1 & s_i^{choice} \neq s_i^{criterion} \end{cases}$$

    end
end

For simplicity, we have omitted some housekeeping details when generating model spaces. For example, it is necessary to supplement the likelihood mappings with contingencies that fall outside isomorphisms. In the current example, one has to add Dirichlet parameters mapping from the *no choice* to a *no feedback* outcome. This is straightforward to implement automatically by ensuring every column of the likelihood tensor has at least one entry. The ensuing model space produces rules of the form:

"If *context* is *context state* then the *criterion state* is the correct *choice*":

In this instance, rule 65 was the rule used to generate feedback:

If *centre* is *red* then the *left* state is the correct *choice*:
if *centre* is *green* then the *centre* state is the correct *choice*:
if *centre* is *blue* then the *right* state is the correct *choice*:

In this form, rule discovery can be seen as probabilistic inductive logic programming (PILP), (Lake et al., 2015; Riguzzi et al., 2014). Probabilistic program induction logic combines inference with first-order logic to learn probabilistic logic programs from data. This entails learning the structure (rules) and parameters (probabilities) of a program, simultaneously or as separate subproblems. Here, the expected information gain over models renders structure learning as efficient as possible, while the expected information gain over parameters ensures efficient parameter learning.

Above, we have considered the likelihood mapping to a modality that underwrites relatively sparse, uninformative feedback. This kind of feedback is typically encountered in reward learning paradigms (where correct and incorrect can be associated with reward and punishment). Crucially, sparse rewards are only delivered when the 'stars are aligned'. On this view, the rules correspond to different ways in which the stars (i.e., states) could be aligned (i.e., in an isomorphic fashion). It is interesting to consider the equivalent generation of model spaces for more informative modalities. For example, one could have used the isomorphism between the levels of the *where* factor and the locations of the three balls—in conjunction with the isomorphism between the location and outcome colours—to specify the likelihood mapping to the first (*visual*) modality. In this instance, the *location* factor would correspond to the context state that specified which of the three criterion states generates outcomes. Here, there is only one model that satisfies these isomorphic constraints, and therefore no need for model selection. Interestingly, this kind of context sensitive model structure is ubiquitous, in the sense that it rests upon a factorisation into 'what' (i.e., *criterion*) and 'where' (i.e., *context*) resulting in the ability to generate the answer to "what would I see if I looked over there?" (Ungerleider, 1994).

Clearly, committing to the above (isomorphic) rules does not offer a universal way of specifying a model space. For example, we could consider the context state to be specified by another context state leading to 3rd order hypotheses. Or, we could have added a further constraint that each context state specifies a unique criterion state, leading to 3×3! = 18 possible rules. However, this would preclude simple (first-order) rules; i.e., a fixed criterion state. From the perspective of the contextual rules, a first-order rule simply means the criterion state is specified when the context state is in the first level OR the second OR the third, and so on. For example:

If *centre* is *red OR green OR blue* then the *left* state is the correct *choice*:

One could also consider constraints afforded by multiple choice factors to increase the size of the model space, if deemed plausible *a priori*. For example, we could have considered rules of the form in which a correct response can only be obtained when looking at the start location:

If *centre* is *red* AND *location* is *start*, then the *left* state is the correct *choice*:
If *centre* is *green* AND *location* is *start*, then …

In summary, under the hypothesis that correct feedback depends upon context-sensitive isomorphisms, the three-ball paradigm offers 81 possible likelihood mappings between latent states and a feedback modality. This set of likelihood models contains the true rule, which has the form: if the state of some contextual factor is this, then the state of the implied criterion factor determines the state of the control factor that satisfies the rule. In this instance, the colour of the central ball determines the location of the ball whose colour corresponds to the correct choice. By construction, this is a relatively simple rule that involve objects having different attributes, such as colour and position. In other words, we did not have to worry about isomorphisms between the location, pose or colour of objects; however, in principle, similar isomorphic constraints can be applied to more expressive generative models. In what follows, the ensuing priors over models are used to simulate artificial reasoning in which an agent has to discover the rule.

## ARTIFICIAL REASONING

This section repeats the simulation in Figure 3 but starting with prior beliefs—about the likelihood model that entails the rule—based on the Bayesian model average of the above rules. Figure 4 shows the results of 64 trials, each comprising six time points (i.e., five moves), starting with Dirichlet priors over the *feedback* likelihood based upon the model average in Equation (8), normalised to a sum of one. Initially, policies—i.e., combinations of eye movement and choice—are evaluated with uninformative prior preferences based upon their expected free energy. This means that visual search behaviour is driven to resolve uncertainty about the colours of the balls at each location, and which of the 81 rules are in play. This means the agent looks at every location and chooses different colours to see whether the outcome is correct or incorrect. As trials progress, the agent accumulates Dirichlet counts in the likelihood mapping, effectively accumulating evidence that speaks to the contingencies mapping from latent states to outcomes. In turn, these are used to evaluate the expected information gain over models using Bayesian model reduction, starting from the initial or full priors.

As evidence accumulates, certain hypotheses are rejected, and the remaining plausible rules are successively compared and eliminated. When the agent is sufficiently confident that the most likely model supervenes over all others, it applies Occam's razor: see Equation (13). In other words, it replaces its posterior beliefs about the likelihood mapping with the mapping selected on the basis of Bayesian model reduction. At this point, informative preferences are adopted and the agent plans, with greater confidence, two steps into the future. This point can be regarded as an 'aha moment' or the dénouement of some data gathering that licences a commitment to a particular hypothesis. This transition is illustrated in Figure 4 in terms of the confidence with which the agent executes its policies. Here, confidence is simply the precision or negative entropy of posterior beliefs over policies. Interestingly, this precision has been used as an explanation of dopaminergic discharges in human neurophysiology (Friston et al., 2014; Schwartenbeck et al., 2015). A marked change in behaviour accompanies the implicit rule discovery or insight. The second panel of Figure 4 demonstrates this in terms of outcomes. During epistemic foraging—to solicit the most informative outcomes—the agent almost makes a choice on every move, which is either correct or incorrect. Both outcomes inform posterior beliefs over rules or hypotheses. However, once uncertainty about the rule has been resolved, the agent secures the maximum number of correct feedbacks, and completely avoids the aversive incorrect outcomes. Note that the maximum number of correct outcomes is either three or four, depending upon whether the central colour is green or not: if the central colour is green, the agent knows to choose green on the next move; otherwise, it has to spend an additional epistemic move looking at the criterion location.

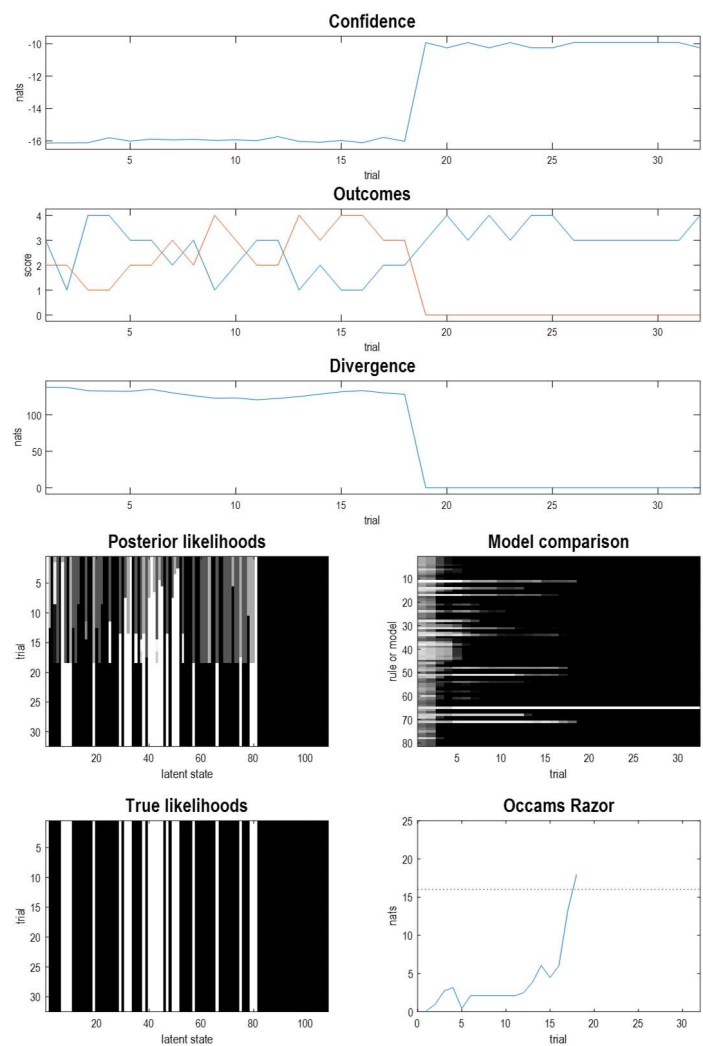

**Figure 4: active reasoning**. This figure reports belief updating and subsequent model selection over 32 trials of the game reported in Figure 3. Each trial starts with randomly selected ball colours, which serve as instructional or contextual cues that enable the agent to resolve uncertainty about the nature of the game. In particular, the agent starts with (81) hypotheses about the likelihood mapping between latent states and the *feedback* modality and accumulates evidence for or against these hypotheses until one is selected (on trial 18). The upper panel reports the confidence the agent has about its responses (i.e., the negative entropy or precision of beliefs over policies). The second panel shows the outcomes. Note that prior to model selection, the agent chooses as many outcomes as it can; irrespective of whether they are *correct* or *incorrect* (blue and red lines, respectively). However, after it has committed to a particular model (here, the correct model, 65) it avoids soliciting surprising or aversive outcomes. It can do this because it now has a precise knowledge of the rule, as shown in the lower panels. The posterior and true likelihoods represent a slice of the likelihood tensor reporting *correct* outcomes over all combinations of latent states. Following model selection, the posterior likelihoods converge on the true likelihoods and the divergence in the third panel falls to 0. At this point, the distribution over models (model comparison) becomes definitive, as reflected in a threshold crossing of Occam's razor, scoring the log odds ratio of the winning model being more likely than the remainder.

Artificial reasoning

During evidence accumulation in the first 18 trials, posterior beliefs about the likelihood mapping remain far from the mapping ultimately selected; although there is a slow reduction, on average, as Dirichlet counts are accumulated. The accumulation of these Dirichlet counts is summarised in the lower panels in terms of a slice through the likelihood tensor; namely, the subtensor mapping from all combinations of latent states to the correct outcome in Equation (14). As time proceeds, the implicit probability distribution becomes increasingly precise until the (true) mapping is selected. At this point, the posterior likelihoods and true likelihoods coincide and the KL divergence in the third panel falls to zero. The panel entitled "model comparison" shows the probability distribution over models based upon Bayesian model reduction; i.e., the likelihood of each model or rule based upon the current posterior, following Equation (9). In this instance, the true rule is number 65 and becomes increasingly predominant as evidence is accumulated. This is reflected in the graph entitled Occam's razor, reporting the log odds ratio in Equation (12). At trial 18, this exceeds a threshold of 16 nats and Occam's razor is implemented. This concludes an illustration of active reasoning, in which action selection or choice behaviour is intrinsically motivated by the expected information gain over both latent states (i.e., the colour of unseen balls) and likelihood mappings (i.e., the rules, contingencies or models that determine whether response is correct or not).

Figure 5 summarises the results of repeating the simulation in Figure 4, 64 times using randomly selected initial conditions. The upper panel reports the KL divergence between the posterior likelihood mapping and the true rule. When the true rule is discovered, this divergence falls to zero, with a threshold crossing of Occam's razor at 16 nats (shown in the second panel). The third panel reports performance in terms of a cumulative score; namely, the number of correct minus incorrect outcomes. Note that prior to rule discovery, the performance score tends to drift down. This is because there are more ways of being wrong than right, but both are informative. Histograms reporting the performance and discovery time are shown in the lower panels. These numerical studies suggest that the number of trials needed to discover the rule depend upon the (random) initial conditions of each trial and subsequent epistemic foraging. In this example, the agent can infer the correct rule after just nine trials, in a couple of instances, but sometimes requires more than 40 trials. The mode of discovery time was about 14 trials with a cumulative score of about 150. But, how much of this sample efficiency can be attributed to active reasoning?

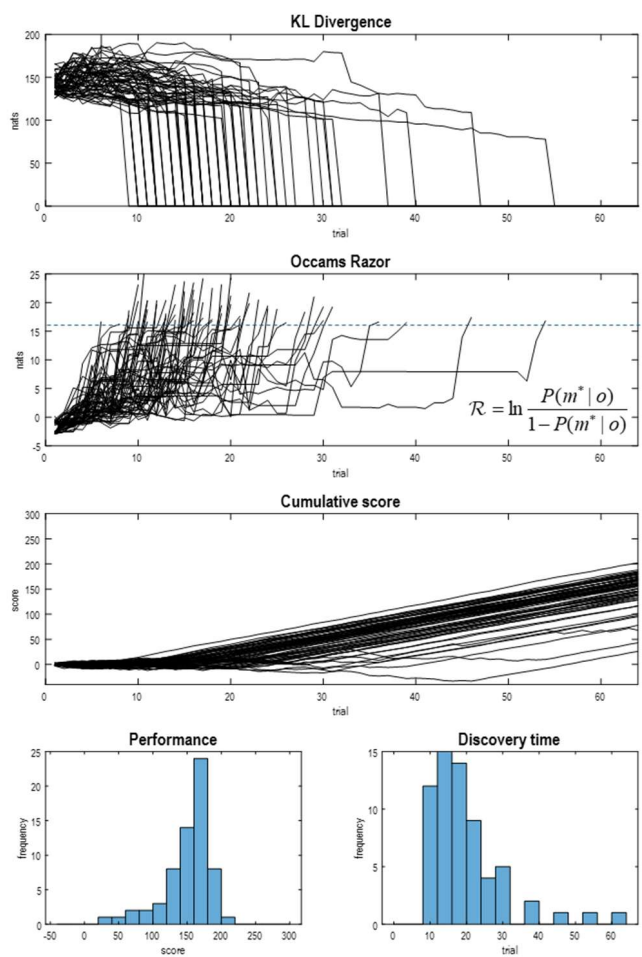

**Figure 5: hypothesis testing**. This figure reports simulations of 64 games, each comprising 32 trials, using the setup described in the previous figure. The upper panel reports the KL divergence between the posterior over the likelihood mapping to the *feedback* modality and the true likelihood mapping. There is, on average, a decrease in divergence until it falls to 0 at the point of Bayesian model selection; i.e., application of Occam's razor. This application rests upon a threshold crossing of the log odds ratio testing for the probability of the most likely model, relative to others. The third panel reports the cumulative score; namely, the difference between *correct* and *incorrect* outcomes. The histograms report the distribution of the cumulative scores and discovery time; i.e., the number of trials before the true model was selected. Note that in one game, Occam's razor was applied prematurely (at trial six) and the agent selected the wrong model. This means the KL divergence increases, and the discovery time is arbitrarily set to 64.

## CONTRIBUTION ANALYSES

Figure 6 provides an intuition about the efficacy of expected information gain in rule discovery or hypothesis testing. The left panels reproduce the evolution of the distribution over models in Figure 4—and accompanying Occam's razor—when the expected information gain includes models, parameters and states. The equivalent results, for exactly the same cues, are provided in the remaining panels when successively removing information gain terms from the expected free energy. When information gain is restricted to parameters and states, active model selection fails to disambiguate between two plausible hypotheses after 32 trials. If we retain just the expected information gain over states, this failure is more pronounced, with two hypotheses being equally plausible after 32 trials. Furthermore, the agent nearly selects the wrong model after 15 trials. In the absence of any expected information gain (i.e., under random behaviour), Occam's razor is nearly zero at trial 32; suggesting that there is no evidence—in the outcomes elicited—to suggest that the winning model is any more likely than the remaining models.

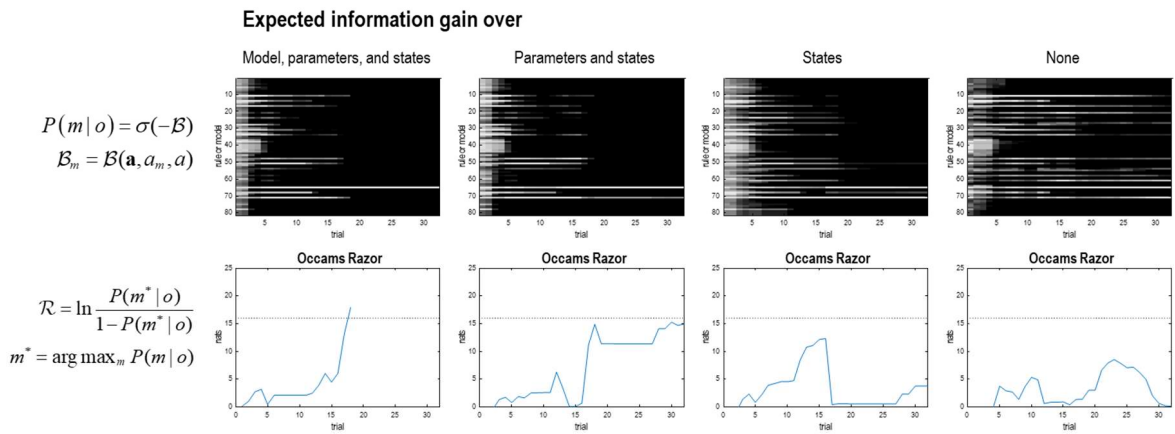

**Figure 6: expected information gain and model selection**. This figure reproduces the results in Figure 4 but in the absence of expected information gain over models, parameters and states. The upper row shows the posterior distribution over models as a function of trials as we remove the expected information gain over models, then parameters, and finally states. The lower panels show the corresponding effect on Occam's razor—and consequent failures to reach threshold—as successive components of expected information gain are removed.

These numerical studies suggest that although expected information gain over parameters is sufficient to eliminate many implausible models, the expected information gain over models becomes important at later stages of evidence accumulation; i.e., resolving uncertainty about a small number of equally plausible models. In what follows, we will take a closer look at the relative contribution of the different sorts of information gain.

We repeated the above numerical study—using the same random initial conditions for each trial—but precluded any expected information gain over models and parameters during the evaluation of expected free energy. The differences in belief updating and performance are summarised in the upper row of Figure 7. The first panel shows the discovery time with and without expected information gain over model parameters. Each dot corresponds to the discovery time—given the same sequence of cues over 64 trials—with and without expected information gain over models and parameters. On average, the discovery time from. With the exception of five instances, the discovery time was universally shorter with expected information gain over models and parameters, reducing from 28.5 to 19.4 trials. A smaller discovery time means that the agent can exploit its insights to secure more correct responses (and avoid incorrect outcomes). This is shown on the upper right with an increase in cumulative performance from 89.2 to 150 points.

An interesting aspect of this information seeking behaviour is that the expected information gain over models and parameters depends upon resolving uncertainty about latent states. This is because the posterior predictive densities—used to evaluate the expected free energy—are more precise when the agent has resolved uncertainty about latent states. The ensuing interdependency can be illustrated by repeating the simulations with and without any expected information gain. The second row of panels in Figure 7 shows the corresponding results in terms of discovery time and performance. In this case, the improvement afforded by information seeking is more marked, with a failure to discover the rule after 64 trials in about half of the simulations (the single failure with expected information gain corresponds to a premature commitment to the wrong model). The corresponding performance fell from an average score of 150 to 51 points. This speaks to a synergistic effect of resolving uncertainty about both states and model parameters.

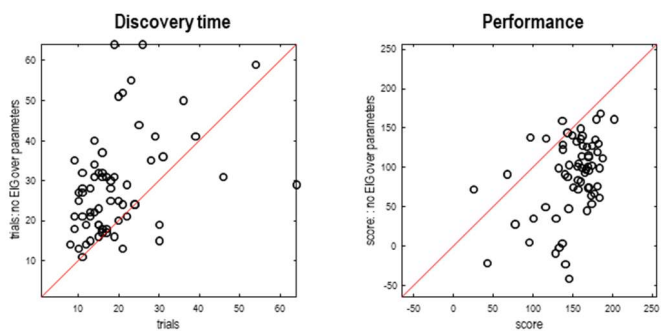

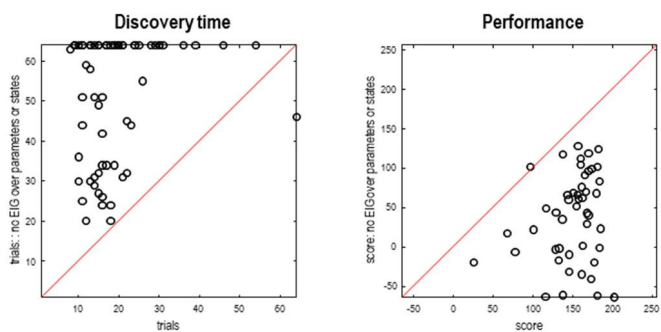

**Figure 7: the importance of being curious**. This figure shows the discovery time and performance when repeating the simulations in Figure 5 with and without expected information gain over models and parameters (upper row). And with and without expected information gain over models, parameters and states (lower row). Each circle corresponds to a game with the same sequence of cues; such that circles above the red lines constitute

worse (respectively, better) results for discovery (respectively, performance) in the absence of expected information gain. A failure to discover the rule after 64 trials is arbitrarily assigned a discovery time of 64.

# CONCLUSION

Active inference provides a formal account of sentient behaviour that is driven purely by maximising the evidence for generative models of exchange between an agent and the sensed world. This is sometimes neatly summarised as self-evidencing (Hohwy, 2016). Technically, this first principles account of self-organisation can be cast as a free energy minimising process—over separable timescales—where variational free energy provides an upper bound on log evidence (a.k.a., marginal likelihood). One useful perspective on this formulation is that one can convert an intractable (exact) Bayesian inference problem into an optimisation problem, through minimising variational free energy (Da Costa et al., 2020; Feynman, 1972; Hinton and Zemel, 1993; Parr and Friston, 2019).

A useful perspective on this optimisation rests upon the fact that the free energy in question is a functional of (approximate) posterior and prior beliefs, where beliefs are used in the sense of Bayesian belief updating and belief propagation (Bagaev and de Vries, 2021; Winn and Bishop, 2005; Yedidia et al., 2005). This means one can minimise variational free energy with respect to the posterior or, crucially, the prior. If one associates different model structures with a set of prior constraints on the parameters of a generative model, one has a straightforward mechanism to absorb structure learning into the minimisation of variational free energy. In particular, one can use Bayesian model reduction to optimise posterior beliefs—and consequent behaviour—under a particular model and optimise the model in and of itself (Friston and Penny, 2011). The current paper addressed the problem of how one should gather data to optimise this kind of structure learning.

In summary, by starting with a set of priors over model parameters, one can start with uniform priors over the implicit model structures and update posterior beliefs as evidence—for or against different models—is accumulated. Furthermore, as illustrated above, one can sample those data that best disambiguate between two or more plausible models. Technically, this just means extending the expected free energy—that underwrites action selection—to cover the expected information gain over models. We have illustrated this using Bayesian model reduction to evaluate a posterior over models—before and after acting under a particular policy—to evaluate the expected information gain, in accord with the principles of optimal experimental design (Lindley, 1956).

The resulting procedures were demonstrated analytically, and with numerical studies, using discrete state-space models; namely, partially observed Markov decision processes. Here, partial observability was inherent in constraints on the sampling of the sensorium (e.g., looking at one place at a time, as an active vision) (Van de Maele et al., 2024). The use of Bayesian model reduction in this setting casts

Artificial reasoningeach alternative model as a reduced version of the Bayesian model average under uniform priors. In other words, each model is defined by removing (combinations of) parameters through suitable shrinkage priors; c.f., (Efron and Morris, 1973; Friston and Penny, 2011; Wipf and Rao, 2007). In the implementation adopted in this work, one simply applies shrinkage priors by setting Dirichlet counts to zero or a small value (e.g., 1/32). This provides a fast and efficient scheme for scoring the likelihood of different models, given just the original priors and a posterior that can be updated. As evidence is accumulated, the ensuing distribution over reduced models becomes increasingly precise, until one model supervenes and can be selected via an application of Occam's razor.

Clearly, there is no guarantee that selected model will account for data that will be encountered in the future. Indeed, if one returns to Figure 4, one of the 64 simulations committed to the wrong model prematurely, after just six trials (the instance of a KL divergence increasing at the point of model selection). In short, there is always a possibility of jumping to conclusions, even with a stringent threshold for applying Occam's razor. This 'jumping to conclusions' is a particularly interesting phenomena, which speaks to the possibility of applying the three-ball paradigm in schizophrenia research, where computational phenotyping suggests that people with schizophrenia have a tendency to jump to conclusions (Averbeck et al., 2011; Joyce et al., 2013; Moutoussis et al., 2011).

As noted in the introduction, this work—and the notion of active reasoning—owes much to the artificial intelligence community's focus on System 2 thinking and, in particular, the ARC-AGI-3 challenge. However, active reasoning, as described in this paper, only addresses a small part of this challenge. First, we have assumed that the agent knows the generative model under which the (unknown) rule is implemented; as opposed to having to learn the generative model from sampling some unsegmented sensory input, such as a pixel array. Furthermore, it has reduced the problem of rule discovery, or abstract reasoning, to discovering the form of a likelihood mapping to a rule-defining outcome. This means that the kind of reasoning demonstrated above rests upon learning a generative model of the world—and the consequences of acting upon that world—before rule learning.

However, if the functional form and structure of the generative model have been learned, or are known, then this (factorial) structure can be used to specify a space of lawful likelihood models that instantiate an unknown rule or program. We have introduced this specification in terms of hypothetical likelihood mappings that feature some context-sensitive symmetry and isomorphisms, which may or may not be applicable to any generative model with a suitable factorial structure. Irrespective of the generality of context-sensitive isomorphic constraints, the fact that these constraints inherit from a factorial structure suggests that the rules about which we reason rest upon carving nature at its joints in the form of conditionally independent factors such as 'what' and 'where', or the attributes of 'objects'. In turn, this speaks to the importance of object-centric representations and requisite factorial models in being able to apply the symmetries or constraints that are variously ascribed to rules, symmetries, equivariances and laws.

# APPENDIX

It is important to distinguish between inference under a Bayesian model average and inferring the posterior over models subtending that average. In the *post-hoc* scheme considered in this paper, inference and learning rest upon minimising the variational free energy in (1).

$$F = \mathbb{E}_Q[\ln \underbrace{Q(s_\tau)Q(u_\tau)Q(\mathbf{A})}_{\text{Posterior}} - \ln \underbrace{P(o_\tau | s_\tau, \mathbf{A})}_{\text{Likelihood}} - \ln \underbrace{P(s_\tau | s_{\tau-1}, u_\tau)P(u_\tau)P(\mathbf{A})}_{\text{Prior}}]$$

$$P(\mathbf{A}) = Dir(a)$$
$$a = \mathbb{E}_{P(m)}[a_m]$$
$$P(\mathbf{A} | m) = Dir(a_m)$$

This is distinct from inferring the model *per se*. Inference over models would require minimising a variational free energy that included an approximate posterior over models:

$$\mathbf{F} = \mathbb{E}_Q[\ln \underbrace{Q(s_\tau)Q(u_\tau)Q(\mathbf{A})Q(m)}_{\text{Posterior}} - \ln \underbrace{P(o_\tau | s_\tau, \mathbf{A})}_{\text{Likelihood}} - \ln \underbrace{P(s_\tau | s_{\tau-1}, u_\tau)P(u_\tau)P(\mathbf{A} | m)P(m)}_{\text{Prior}}]$$

However, inference and learning under the prior over models, i.e., $Q(m) = P(m)$, renders the variational free energy in (1) a lower bound approximation to the variational free energy required for inference over models.

$$\mathbf{F} = F + \mathbb{E}_Q[\ln Q(m) - \ln P(m)] - \mathbb{E}_Q[\ln P(\mathbf{A} | m) - \ln P(\mathbf{A})]$$
$$= F - \mathbb{E}_Q[\ln P(\mathbf{A} | m) - \ln P(\mathbf{A})]$$
$$= F - \mathbb{E}_{Q(\mathbf{A})}[\mathbb{E}_{P(m)}[\varphi(a_m)] - \varphi(\mathbb{E}_{P(m)}[a_m])]$$
$$\Rightarrow \mathbf{F} \geq F$$

This rests on the Jensen inequality, $\mathbb{E}[\varphi(a)] \leq \varphi(\mathbb{E}[a])$, given that the digamma function $\psi$ is strictly concave on $(0,\infty)$ and $\varphi(a) = \psi(a) - \psi(a_0)$. On this view, updating the approximate posterior over models—by application of Occam's razor—can be regarded as a discrete (*post-hoc*) approximation to continual structure learning; i.e., inference over models. One might conjecture that accumulating evidence under uninformative model priors—before updating posterior beliefs about hypotheses—ensures the above inequality holds. Practically, a *post hoc* scheme means one does not have to entertain multiple models during active inference and learning.

**Acknowledgements**

KF is supported by funding from the Wellcome Trust (Ref: 226793/Z/22/Z). TP is supported by an NIHR Academic Clinical Fellowship (ref: ACF-2023-13-013).

**Disclosure statement**

The authors have no disclosures or conflict of interest.

**Software availability statement**

The MATLAB routines used to generate the simulations reported here will be made publicly available via the SPM 25 academic software package (https://www.fil.ion.ucl.ac.uk/spm/) and is available on request from the authors pending this release.

Artificial reasoning

Artificial reasoningMazzaglia, P., Verbelen, T., Catal, O., Dhoedt, B., 2022. The Free Energy Principle for Perception and Action: A Deep Learning Perspective. Entropy (Basel) 24, 301.
Moutoussis, M., Bentall, R.P., El-Deredy, W., Dayan, P., 2011. Bayesian modelling of Jumping-to-Conclusions bias in delusional patients. Cogn Neuropsychiatry 16, 422-447.
Olshausen, B.A., Field, D.J., 1996. Emergence of simple-cell receptive field properties by learning a sparse code for natural images. Nature 381, 607-609.
Oudeyer, P.-Y., Kaplan, F., 2007. What is intrinsic motivation? a typology of computational approaches. Frontiers in Neurorobotics 1, 6.
Parr, T., Friston, K., Zeidman, P., 2024. Active Data Selection and Information Seeking. Algorithms 17, 118.
Parr, T., Friston, K.J., 2017. Working memory, attention, and salience in active inference. Scientific reports 7, 14678.
Parr, T., Friston, K.J., 2019. Generalised free energy and active inference. Biological cybernetics 113, 495-513.
Parr, T., Pezzulo, G., Friston, K.J., 2022. Active Inference: The Free Energy Principle in Mind, Brain, and Behavior. MIT Press, Cambridge.
Penny, W.D., 2012. Comparing Dynamic Causal Models using AIC, BIC and Free Energy. Neuroimage 59, 319-330.
Riguzzi, F., Bellodi, E., Zese, R., 2014. A History of Probabilistic Inductive Logic Programming. Frontiers in Robotics and AI Volume 1 - 2014.
Ruffini, G., 2017. An algorithmic information theory of consciousness. Neuroscience of consciousness 2017, nix019.
Sakthivadivel, D.A.R., 2022. Weak Markov Blankets in High-Dimensional, Sparsely-Coupled Random Dynamical Systems, p. arXiv:2207.07620.
Sanchez, E.H., Serrurier, M., Ortner, M., 2019. Learning Disentangled Representations via Mutual Information Estimation, p. arXiv:1912.03915.
Savage, L.J., 1954. The Foundations of Statistics. Wiley, New York.
Schmidhuber, J., 2010. Formal Theory of Creativity, Fun, and Intrinsic Motivation (1990-2010). Ieee Transactions on Autonomous Mental Development 2, 230-247.
Schwartenbeck, P., FitzGerald, T.H., Mathys, C., Dolan, R., Friston, K., 2015. The Dopaminergic Midbrain Encodes the Expected Certainty about Desired Outcomes. Cereb Cortex 25, 3434-3445.
Schwartenbeck, P., Passecker, J., Hauser, T.U., FitzGerald, T.H., Kronbichler, M., Friston, K.J., 2019. Computational mechanisms of curiosity and goal-directed exploration. eLife 8, e41703.
Silver, D., Singh, S., Precup, D., Sutton, R.S., 2021. Reward is enough. Artificial Intelligence 299, 103535.
Tenenbaum, J.B., Kemp, C., Griffiths, T.L., Goodman, N.D., 2011. How to grow a mind: statistics, structure, and abstraction. Science 331, 1279-1285.
Tervo, D.G.R., Tenenbaum, J.B., Gershman, S.J., 2016. Toward the neural implementation of structure learning. Curr Opin Neurobiol 37, 99-105.
Tipping, M.E., 2001. Sparse Bayesian learning and the relevance vector machine. Journal of Machine Learning Research 1, 211-244.
Tomasello, M., 2016. Cultural Learning Redux. Child development 87, 643-653.
Tsividis, P., Loula, J., Burga, J., Arnaud, S., Foss, N., Campero, A., Pouncy, T., Gershman, S.J., Tenenbaum, J.B., 2021. Human-Level Reinforcement Learning through Theory-Based Modeling, Exploration, and Planning. ArXiv abs/2107.12544.
Ungerleider, L., 1994. 'What' and 'where' in the human brain. Current Opinion in Neurobiology 4, 157-165.
33